\documentclass[%
 aip,
 amsmath,amssymb,
preprint,
]{revtex4-1}

\usepackage{graphicx}
\usepackage{dcolumn}
\usepackage{bm}

\usepackage[utf8]{inputenc}
\usepackage[T1]{fontenc}
\usepackage{mathptmx}
\usepackage{etoolbox}
\usepackage{xcolor}
\makeatletter
\def\@email#1#2{%
 \endgroup
 \patchcmd{\titleblock@produce}
  {\frontmatter@RRAPformat}
  {\frontmatter@RRAPformat{\produce@RRAP{*#1\href{mailto:#2}{#2}}}\frontmatter@RRAPformat}
  {}{}
}%
\makeatother
\begin{document}

\preprint{AIP/123-QED}

\title[]{In-situ characterization of optical micro/nano fibers using scattering loss analysis}
\author{Shashank Suman}
\author{Elaganuru Bashaiah}%
\author{Resmi M}
\affiliation{ 
School of Physics, University of Hyderabad, Hyderabad, Telangana, India-500046
}%

\author{Ramachandrarao Yalla}
\email{rrysp@uohyd.ac.in}
\affiliation{ 
	School of Physics, University of Hyderabad, Hyderabad, Telangana, India-500046
}%

\date{\today}

\begin{abstract}
We experimentally demonstrate the in-situ characterization of optical micro/nano fibers (MNFs).The MNF (test fiber, TF) is positioned on a microfiber (probe fiber, PF) and simulated for the scattering loss at various PF and TF diameters. The TF is fabricated using chemical etching technique. The PF is a conventional single-mode fiber with an outer diameter of 125 $\mu$m. We measure the scattering loss along the TF axis at various positions i.e. diameters by mounting it on the PF. The diameter profile of the TF is inferred from the measured scattering loss and correlated with its surface morphology measurement.  This work demonstrates an effective, low-cost, and non-destructive method for in-situ characterization of fabricated micro/nano fibers (OMNFs). It can detect and determine the irregularities on the surface of OMNFs. It can also be used to quantify the local evanescent field. Detecting such local points can improve studies that are carried out using these fields in various sensing and related study domains. It is simple to implement and can be accessed by all domains of researchers.
\end{abstract}
\maketitle

\section{Introduction}
With the advent of quantum technologies, the necessity to explore properties and possible applications of an optical fiber with micro/nano dimensions attracts great interest \cite{01,02,03,04,05,06,07,08,09,010,011,012,013,014,015,016,017,018,019,020,021,022}. It offers a compact and feasible option for various applications in the technical arena \cite{01,02,03,04,05,06,07,08,09,010,011,012,013,014,015,016,017,018,019,020,021,022}, such as quantum optics \cite{01}, fiber-coupled quantum light sources \cite{02,03,04,05}, photonic devices \cite{06}, sensing \cite{07,08}, health, environmental tracking and medicine \cite{09,010,011,012,013,014,015,016,017,018,019}, communication and networking \cite{020,021,022}. In the above-mentioned research fields, determining the diameter of optical micro/nano fibers (MNFs) is vital. The scanning electron microscope (SEM) and the atomic force microscope (AFM) are commonly existing methods. 

The measurement accuracy using the SEM is in a few nanometers, additionally providing details on surface morphology, but the method itself is destructive. The AFM images surface morphology with very high precision, but it is difficult to determine the precise diameter of the free-hanging fiber. These equipments are expensive, technically challenging to handle, and require samples to be mounted on a substrate, which is not suitable for free-hanging samples like MNFs. Thus, an in-situ, reliable, and non-destructive method for determining MNF diameters locally is crucial. 

Several non-destructive approaches have been proposed and reported \cite{023,024,025,026,027,028,029,030,031,032,035}. Examples would include complex diffraction patterns \cite{023}, composite photonic crystal cavity \cite{024}, harmonic generation methods \cite{025}, modal interference techniques \cite{026}, stress-strain analysis \cite{027}, whispering gallery modes \cite{028} and different types of scattering methods \cite{029,030,031,032,035}. In the complex diffraction pattern experiment, the reported accuracy was $\pm$50 nm \cite{023}, but it required the surface to be smooth and could not account for surface variation. The reported resolution using the composite photonic crystal cavity method is $\pm$10 nm \cite{024}, but it requires additional fabrication of photonic crystal structures. The harmonic generation technique accuracy is $\pm$2 nm \cite{025}. However, it was also noticed that this method has induced some irreversible changes in the fiber, including a change in phase-matching wavelength due to an inherently weak nonlinear response \cite{033}. The modal interference method produces a resolution of $\pm$6 nm \cite{026}, but it comes with the limitation that only an allowed higher-order mode family can provide beat frequencies that can be measured. The stress-strain analysis method is accurate up to $\pm$8 nm \cite{027}, but the tensile force limits it and can also damage the sample. The whispering gallery mode method was accurate up to $\pm$2 nm \cite{028}, but it needs both test fiber (TF) and probe fibers (PF) to be fabricated for maximum optimized output.

Furthermore, different types of scattering methods are also reported with different accuracy \cite{029,030,031,032}. The accuracies of $ \pm$10 nm and $\pm$3 nm are experimentally demonstrated for Brillouin and Rayleigh scattering, respectively. One of the scattering methods also reported an accuracy of $\pm$0.7 nm \cite{032}. The scattering method using a microfiber tip reported an accuracy of 9.8 nm \cite{035}. Although these proposed methods exist, there is still scope for developing new methods to measure MNFs optically. In general, the above-mentioned methods require complex structures and procedures. Hence, a simple and cost-effective method is required.

In this paper, we conduct an experimental demonstration to showcase the in-situ characterization of MNFs. The TF is positioned on the PF and simulated for the scattering loss at different diameters of both the PF and TF. The fabrication of the TF employs chemical etching technique. The PF utilized in this experiment is a conventional single-mode fiber (SMF) with an outer diameter of 125 $\mu$m. By mounting the TF onto the PF, we measure the scattering loss at different positions along the TF axis, thus enabling us to infer the diameter profile of the TF. This inference was made by analyzing the measured scattering loss and correlating it with the surface morphology measurement of TFs. The experimental results agree with the simulation-predicted results, with the error being within acceptable limits.
\begin{figure}[h!]
	\centering
	\includegraphics[width= 12 cm]{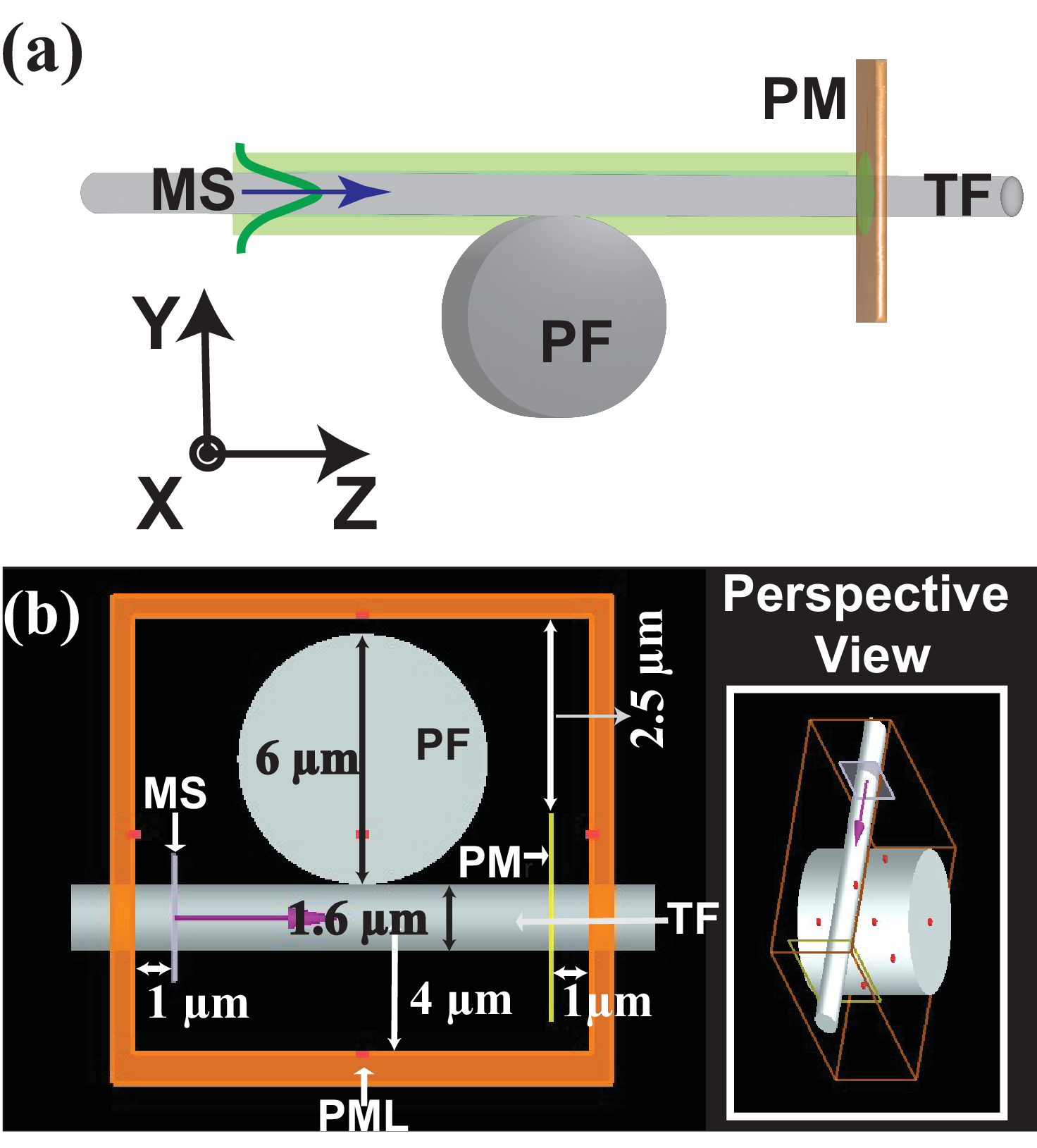}
	\caption{(a) A conceptual diagram of the idea. PF, TF, MS, PM, and PML denote probe fiber, test fiber, mode source, power monitor, and perfectly matched layer, respectively. (b) The left panel shows the cross-section of the corresponding finite-difference time-domain simulation layout. The right panel shows the perspective view.}
	\label{fig1}
\end{figure} 
\section{Simulations}
\subsection{Procedures}	
Figure \ref{fig1} (a) shows a conceptual idea of the proposed method. Using the three-dimensional finite-difference time-domain (FDTD package, Ansys) method, firstly we perform a detailed investigation of in-situ diameter measurement of MNFs by systematically tracking the scattering loss for various diameters of TF/PF. The MNF is used as the TF. The other fiber with a thicker diameter is used as the PF. A mode source (MS) is used to excite guided modes of the TF. We only excite fundamental mode in the TF through MS during theoretical simulations. The light propagates from the MS in the $z$-direction i.e., along the TF axis. A power monitor (PM) is placed to record the transmitted power through guided modes of the TF. The PF is placed along the $x$-axis, which is perpendicular to the TF. As the diameter of the TF is varied, the PF is moved along the $y$-direction to keep fibers in contact, allowing us to study the effects on scattering loss. In a similar fashion, PF diameter is also varied. 
\begin{figure}[h]
	\centering
	\includegraphics[width= 14 cm]{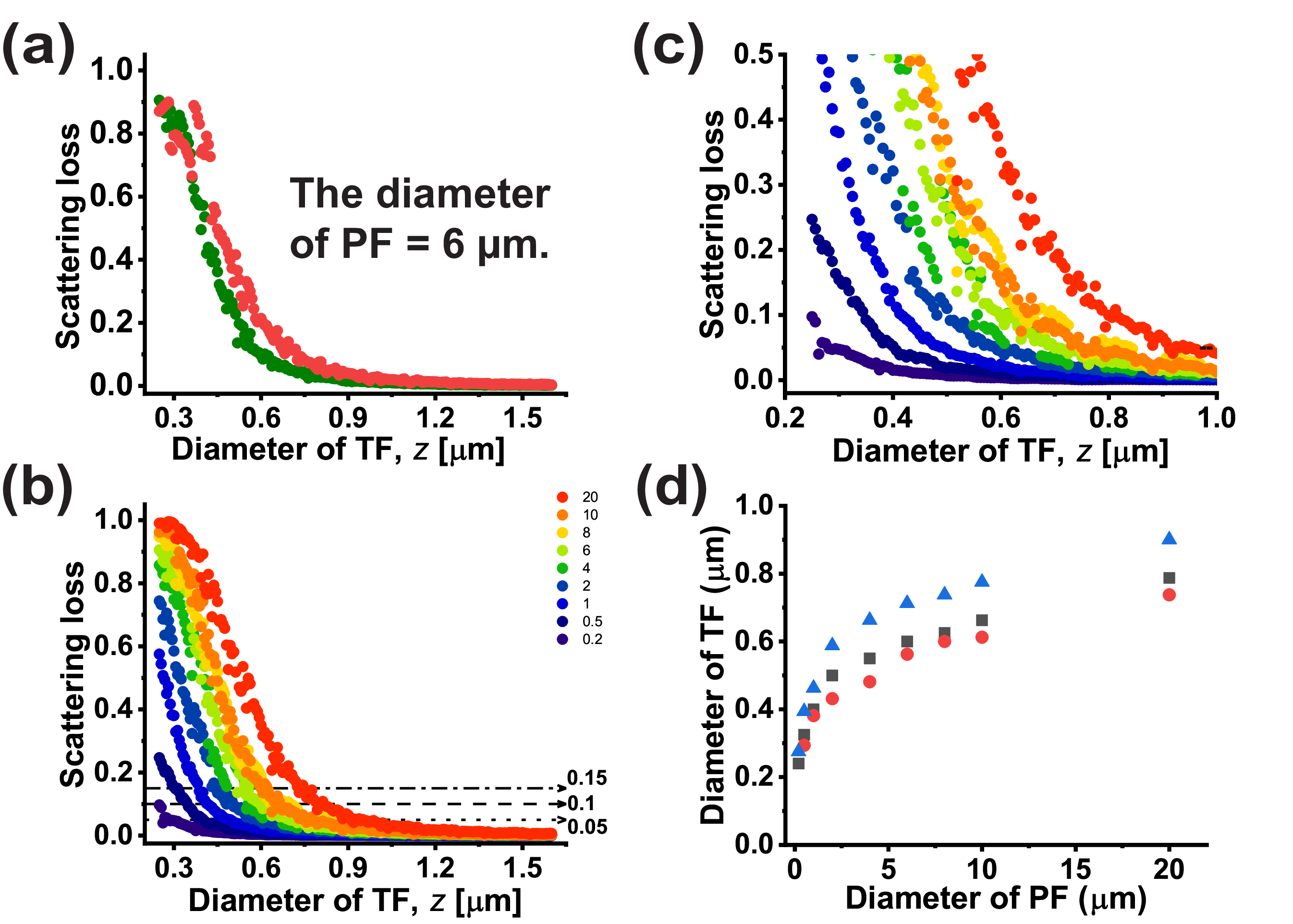}
	\caption{(a) Scattering loss as a function of the test fiber (TF) diameter when the diameter of the PF is 6 $\mu$m. Green and red solid circles correspond to a wavelength of 532 nm and 633 nm, respectively. (b) Scattering loss variation at the wavelength of 532 nm for various PF diameters. (c) An enlarged view depicting the region of interest of (b) in detail. (d) The diameters of PF versus TF for a particular scattering loss. The blue triangles, gray squares, and red solid circles denote the scattering losses of 5$\%$, 10$\%$, and 15$\%$, respectively. }\label{fig2}
\end{figure}
The FDTD simulation layout is shown in Fig. \ref{fig1} (b). We use a silica cylinder with a diameter of 6 $\mu$m as the PF. The diameter of the other silica cylinder is set at 1.6 $\mu$m for the TF. The diameter range for the TF (0.25 $\mu$m to 1.6 $\mu$m) was selected considering the experimental feasibility in our experiments. The PF was positioned perpendicularly on the surface of the TF, as described in Fig. \ref{fig1} (a). The length of TF is 12 $\mu$m, so that it is well outside the PML enclosure. Considering the experimental availability, the MS is set at different wavelengths ($\lambda$) of 535 and 633 nm. The PM is a frequency domain field monitor for recording transmission as a function of TF diameter. The distance between PM and MS is 9 $\mu$m to avoid back reflections. The refractive index of the surrounding medium is set to 1. In the right panel of Fig. \ref{fig1}. (b), we can see a perspective view in 3D. Under the same conditions, the simulation is performed for different diameter values of PF.

\subsection{Results}
Figure \ref{fig2} (a) shows results obtained with the PF diameter of 6 $\mu$m. The plot shows the scattering loss as a function of the TF (MNF) diameter for both wavelengths. Red and green circles are for 633 and 535 nm, respectively. One can readily see that as the diameter of TF increases, there is a gradual decrease in scattering loss.

The simulations are performed with various PF diameters ranging from 0.2 $\mu$m to 20 $\mu$m. The results are summarized in Fig. \ref{fig2} (b). The left and right-most plots correspond to PF of diameters 0.2 $\mu$m and 20 $\mu$m, respectively. One can readily infer that with the increase in PF diameter, the scattering loss increases, provided all the other factors are kept constant.  In Fig. \ref{fig2} (c), an enlarged view of the region in the box is shown. To choose the PF diameter for the experiment, three specific points corresponding to scattering loss of  15$\%$, 10$\%$, and 5$\%$ for all TF/PF diameters are chosen considering the experimental ambiguity. Figure \ref{fig2} (d) shows the diameter of PF versus TF for various scattering loss values. The blue triangles, gray squares, and red solid circles represent 5$\%$, 10$\%$, and 15$\%$, respectively. The trend for each scattering loss remains the same and unique to PF-TF diameter. It shows a significant drop when the diameter of the PF drops below 2 $\mu$m. Note that the relative gap between the selected scattering loss data points increases for thicker PF diameters for the given TF diameter range. 

The single-mode condition is defined by the parameter $V$= $ka\sqrt{n_{1}^{2} - n_{2}^{2}}$$<$ 2.405, where $a$ is the fiber radius, $k$= $2\pi$/$\lambda$, $\lambda$ is wavelength of the light, and $n_{1}$ and $n_{2}$ are refractive indices of the core and clad, respectively \cite{034}.	For the wavelength of 532 nm which is used in our experiments, the maximum diameter is $\sim$0.39 $\mu$m. It allows only the TF's fundamental mode (HE$_{11}$). For the fiber diameter thicker than $\sim$0.39 $\mu$m, higher order modes are excited in the TF, which are TE$_{01}$, TM$_{01}$, and HE$_{21}$\cite{034}. A higher and lower effective refractive index implies that most of the power is in propagating mode inside the core and cladding, respectively. The mode experiencing a lower index interacts more with the PF due to the larger overlap. For 633 nm, the single mode condition is  $\sim$0.46 $\mu$m. Therefore, the evanescent tail is longer for the red light source compared to the green.

	As seen in Fig. \ref{fig2} (a), one can readily observe continuous gradual increases in scattering loss. This leads to accurately determining the TF diameter values. Simulation results are consistent with the mode structure for both sources \cite{034}. For the red source, kinks are shifted to the left, and additional anomalies are observed. This is due to a higher intensity in the evanescent field for a given diameter of the TF, leading to a higher scattering loss. Details about mode excitation and power in the evanescent region can be found in Ref. \cite{034}. As seen in Fig. \ref{fig2} (b), PFs with a higher diameter show variations other than expected kinks. A major shift in scattering loss is observed in a couple of PF diameters. This behavior may be attributed to resonant coupling and the beating of modes. The general behavior of our interest i.e. the nature of scattering loss and its dependence on the diameter of the TF and PF still remains the same. For thicker PF diameters than thinner ones, gradual changes in the scattering loss and minimal deviations are clearly observed. This produces better resolution in determining the TF diameter as clearly seen in Fig. \ref{fig2} (c). This is due to the thicker diameter; the overlap between the PF and evanescent region is larger. One can readily see that as the diameter of the PF increases, the gap between scattering losses of 5$\%$ and 10$\%$ increases. This suggests that choosing the PF diameter thicker is better for the experiment. It also suggests that the to-be-determined range of the TF diameter increases. A similar increase in the gap is also observed between 10$\%$ and 15$\%$. 

\section{Experimental}
\subsection{Procedures}
\begin{figure}[h]
	\centering
	\includegraphics[width= 15 cm]{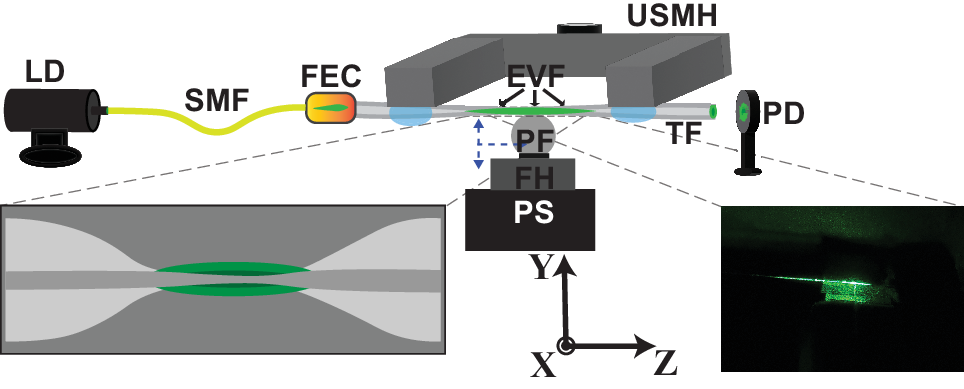}
	\caption[6pt]{The experimental setup used for the measurements. LD,  FEC, PS, USMH, FH, PD, PF, TF, SMF,  and EVF denote laser diode, fiber end connector, precision stage, U-shaped metal holder, probe fiber holder probe fiber, photo-diode, probe fiber, test fiber, single mode fiber,  and evanescent field, respectively. The left and right panels in the inset show an EVF region and scattering observed when the TF comes in contact with the PF, respectively. 		
	}\label{fig3}
\end{figure}
Next, we proceed with experiments based on the simulation results. A conventional SMF (1550B-HP, Coherent) is stripped using a stripping tool (FTS4, Thorlabs) and fixed on a U-shaped metal holder (USMH). Alternatively, acetone is used to remove the outer plastic jacket. The chemical method reduces surface roughness compared to the mechanical method. The TF was fabricated using a modified chemical etching technique \cite{036}, which involves two steps. The different concentrations of hydrofluoric (HF) acid, 40$\%$ and 24$\%$ are used in the first and second steps, respectively. The first step is to etch most of the cladding region rapidly. The second step is slowed down to ensure control over the etching of the core region. The detailed procedures of the process will be reported elsewhere. After the fabrication of the TF, the USMH is mounted on a stationary metal board for mechanical stability.

Figure \ref{fig3} shows a schematic of the experimental setup. Laser diodes (LD) at wavelengths of 532 nm (DL-G-5, Holmarc) and 633 nm (HNLS008L-EC, Thorlabs) were used for the experiment. The laser light is guided using mirrors (BBI-E02, Thorlabs) and input power is controlled by a variable neutral density filter (HO-VND-N50, Holmarc). The ends of the TF are cleaved perpendicularly using a fiber cleaver (XL411, Thorlabs) and inserted in the fiber end connector (FEC) (B30126A9, Thorlabs). The light is guided into the SMF (SMF600, Thorlabs) through a fiber collimator (FC) (P3-405B-FC-1, Thorlabs), and modes are excited in the fabricated TF. As the light passes through the TF, the intensity of light in the evanescent region increases with a decrease in the diameter, thereby increasing the loss due to the scattering by the PF. To avoid complications due to multi-mode in TF, we couple the light into TF through a SMF which filters other modes, thereby allowing only single mode, i.e., fundamental mode to be excited in the TF.Considering the experimental ambiguities and based on the simulation results, the PF used in the experiment is a conventional SMF with an outer diameter of 125 $\mu$m. The PF is fixed on a micro-positioning stage (PS) (TS-65, Holmarc) for translating in the $y$ and $z$-directions. To avoid any damage from the PF on the TF, we minimize the contact time. Once we take the reading from a particular position, the TF is unmounted from the PF and the moved along z-direction in next position, before mounting again for measurement.

The left inset of Fig. \ref{fig3} shows an enlarged view of the waist region. The PF is placed such that it just touches the TF.  The scattering loss with the position of the PF along the TF axis is tracked using a photodiode (PD) (B120VC, Thorlabs). At a given position, the contact is maintained for a period of 20 s, and the transmission data is acquired from a PC connected to a console for an optical power meter (COP) (PM100D, Thorlabs). We define $z$= 0 as the center of the TF. The PF is translated along the $z$-axis from -2.5 mm to 2.5 mm for measuring a diameter profile.	The experiment was performed on an optical bench in a clean booth to avoid contamination. An enlarged view corresponding to the scattering is shown on the right side of the inset of Fig. \ref{fig3}.
After the measurements, the sample was carefully fixed onto a metal holder. Note that the evanescent region was marked on the metal plate. The metal holders were sputtered and imaged using a field emission scanning electron microscope (FESEM) for diameter measurements of TFs.

\subsection{Results}
\begin{figure}[h]
	\centering
	\includegraphics[width= 14cm]{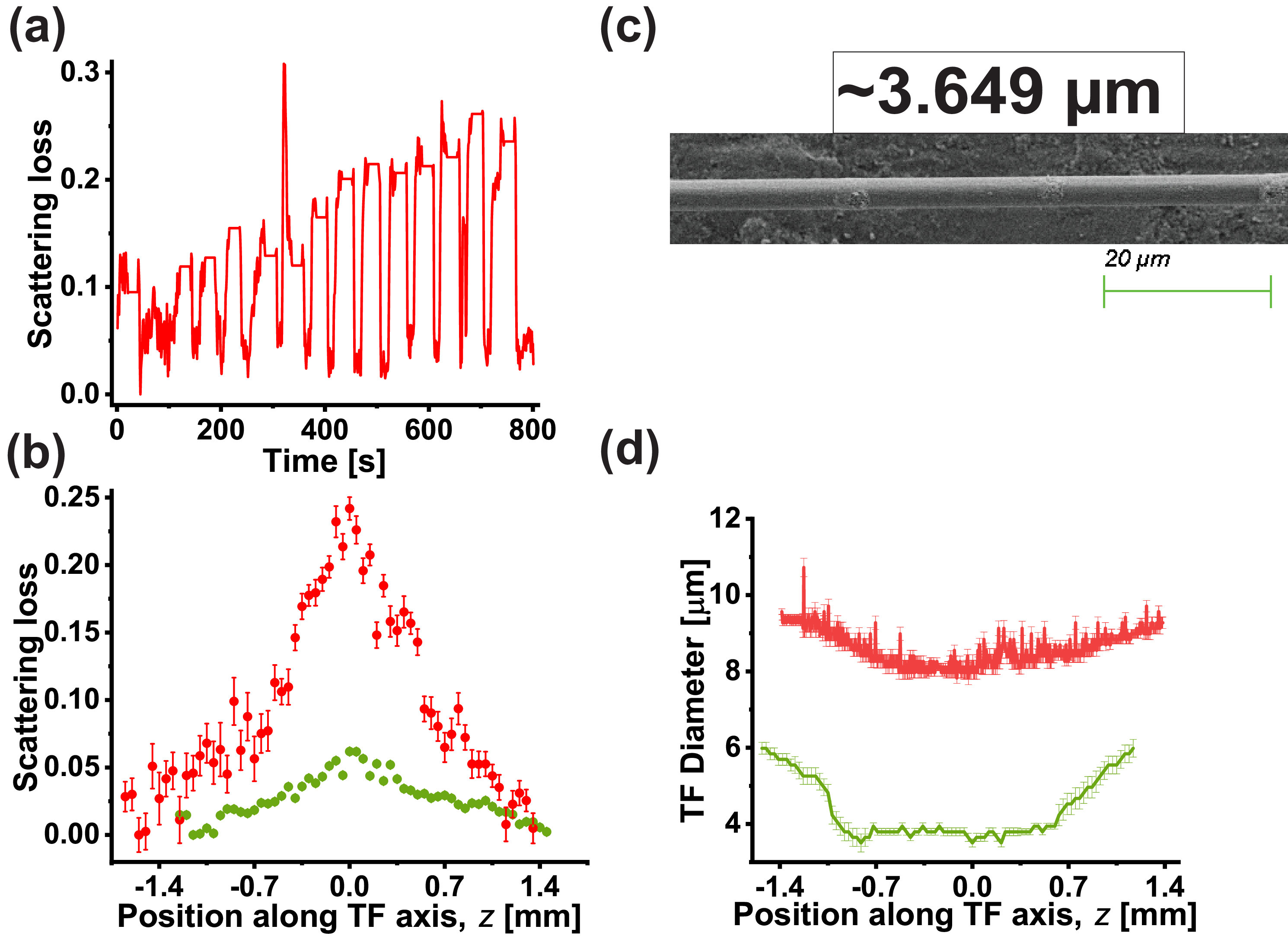}
	\caption{(a) A typical plot for the scattering loss as a function of time.  (b) A typical summary for scattering loss as a function of position along the test fiber (TF). (c) A typical field emission scanning electron microscope image of the TF. (d) A typical summary of the measured diameter profile along the TF axis as a function of TF length. The red and green colored data points are for the red and green sources, respectively.}\label{fig4}
\end{figure}

Figure \ref{fig4} (a) shows a typical scattering loss against time. The PF is moved towards the right from the left. It is clearly observed that as the PF is moved along the axis of the TF, there is a significant increase in the scattering loss. 

Figure \ref{fig4} (b) shows a typical summary of scattering loss against position along the TF axis. Green and red solid circles are for 532 nm and 633 nm, respectively. The transmission readings reported here are normalized to the initial power received by the PD i.e. non-contact.  One can readily see that scattering loss increases and then decreases for both wavelengths, as predicted by the simulations. Note that the trend of rise and fall in scattering loss with respect to position is consistent with both laser sources. The maximum scattering losses recorded in Fig. \ref{fig4} (b) are $\sim$ 7$\%$ and $\sim$ 24$\%$ for green and red sources, respectively. 
Figure \ref{fig4} (c) shows a typical FESEM image of the TF. The diameter of the TF in the image is found to be 3.649 $\mu$m. In a similar fashion, images are collected for each location within the marked region. The summary of the results is shown in Fig. \ref{fig4} (d). One can readily observe the diameter profile of the TF, which corresponds to the scattering loss profile of the TF in Fig. \ref{fig4} (b).   
\begin{figure}[h]
	\centering
	\includegraphics[width= 14 cm]{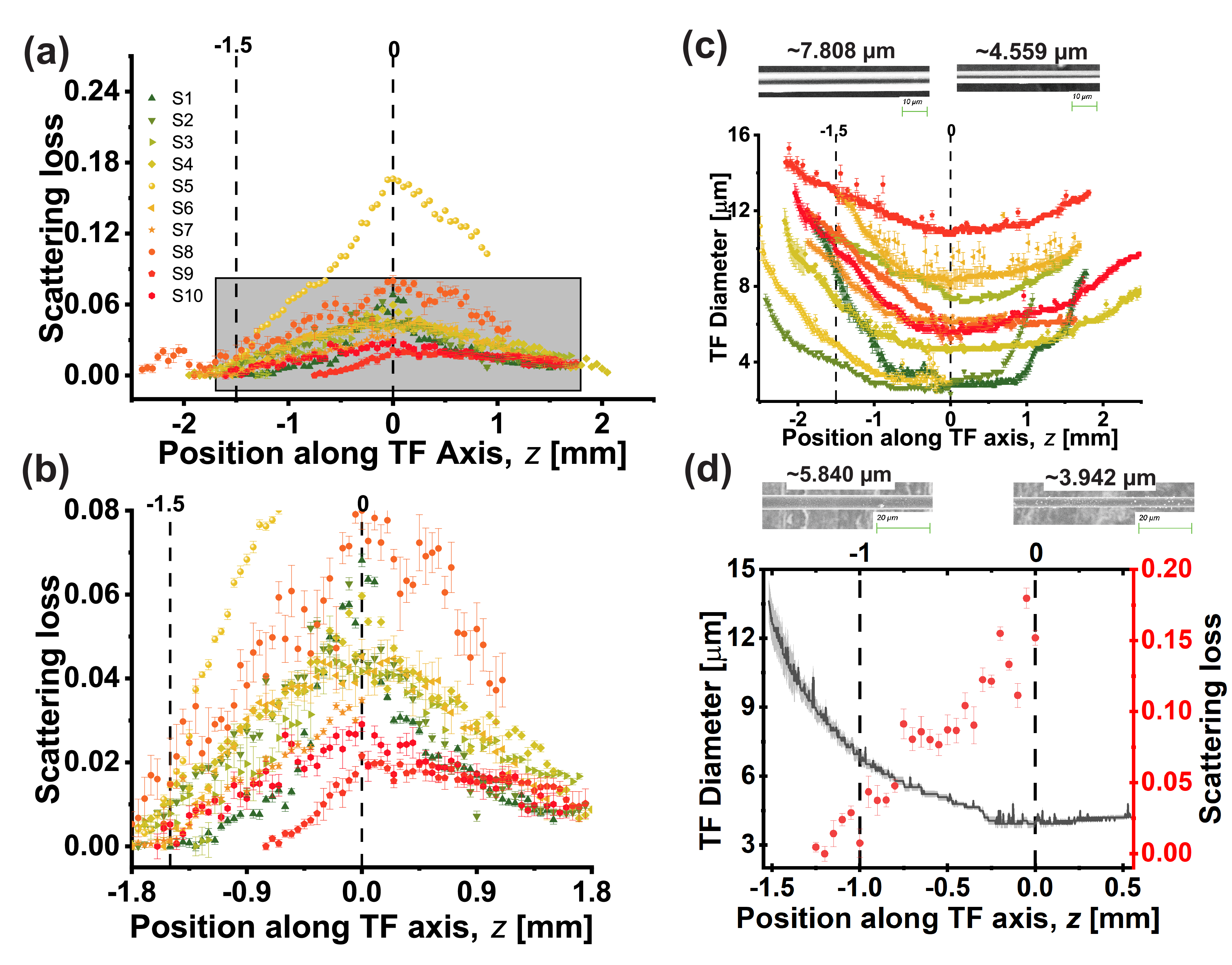}
	\caption{(a) The summary for the scattering loss for various test fibers (TF) for green laser. (b) An enlarged portion of (a). (c) The corresponding diameter profiles are measured using the field emission scanning electron microscope (FESEM). The inset shows typical images. (d) The summary for the scattering loss for various TFs for red laser is shown by red dots. The corresponding diameter profiles are measured using the FESEM is shown as line graph. The inset shows typical images. }\label{fig5}
\end{figure}

Figure \ref{fig5} summarizes the measured results for several TFs. In Fig. \ref{fig5} (a) for the green laser source, the scattering losses versus positions along the TFs axes are plotted. Different colors and symbols represent different TFs. The general trend of increasing and then decreasing remains the same for all TFs. The measured results show a considerable scattering loss, ranging from $\sim$2$\%$ to $\sim$18$\%$ as predicted by the simulations. The corresponding measured diameter profiles of the TFs are shown in Fig. \ref{fig5} (c). The waist diameter of the TF varies from $\sim$2 $\mu$m to $\sim$11 $\mu$m. The typical waist diameter of the TF is shown in the inset. The left panel of the inset corresponds to the measurement at -1.5 mm. In Fig. \ref{fig5} (d) for the red laser source, the scattering losses versus positions along the TFs axes are plotted. The general trend of increasing and then decreasing remains the same for all TFs. The measured results show a considerable scattering loss of $\sim$20$\%$ as predicted by the simulations. The corresponding measured diameter profiles of the TFs are also shown in Fig. \ref{fig5} (d) as line graph. The waist diameter of the TF is $\sim$4 $\mu$m. The typical waist diameter of the TF is shown in the inset. The left panel of the inset corresponds to the measurement at -1 mm. 

As seen in Fig. \ref{fig4} (a), when the PF is moved in the $z$-direction from -1.5 mm to 0 mm, the overlap between the evanescent field in TF and the PF increases. As we unmount TF, we see a dip in the scattering loss as expected. A few kinks are seen at different times. This may be due to the sudden change in the TF diameter and mounting conditions. As seen in Fig. \ref{fig4} (b) for various points along the TF axis, it is observed that the scattering loss for the longer wavelength (red) is higher than for the shorter wavelength (green) as predicted by the simulations in Fig. \ref{fig2} (a). It should be noted that throughout the experiment, the PF diameter of 125 $\mu$m was kept the same.  The variation in the scattering loss correspond to the variation in the TF diameter as seen in Fig. \ref{fig4} (d). The measured behavior of the scattering loss is consistent with the measured diameter profile. Therefore, it can be used to probe any kind of variation in the diameter of the TF. One can also infer the length of the waist region of the TF. 

As seen in Figs. \ref{fig5} (a) and (d), a clear distinction can be observed that each sample has its own unique scattering loss profile. One can also infer that the length of the waist region varies from sample to sample due to the fabrication ambiguity. All samples show different maxima in scattering loss. Note the intervals between consecutive scattering losses are the same for all the samples. As seen in Figs. \ref{fig5} (b) and (d), one can readily observe that no two samples are replicated during the fabrication as expected for the chemical etching technique. Due to the surface roughness of the TF, the scattering loss readings around these points do not follow the usual trend, and either a sudden dip or a sudden peak is observed. However, it is possible to track them down using a scattering experiment, as shown in Figs. \ref{fig5} (a) and (d).  We observe that jumps can be more distinctively observed when a wavelength of 633 nm is used for probing. The correspondence between Figs. \ref{fig5} (a) and (c) and in Fig. \ref{fig5} (d) are good. We have established that scattering at local point is proportional to the variation of the local diameter in the simulations. We can extend those values for PFs used in the experiment. These can track the variation of the diameter precisely. The absolute value of diameter cannot be obtained, as we do not have a calibration curve. This is the limitation of using the chemically etched micro/nano fibers. However, this method offers an economical and non-destructive solution to quantification of field intensity in the evanescent region.

 Although ambiguities exist in the measured results, they are all consistent with the simulation results as discussed earlier in this section. The resolution of the method is limited by the range of the TF diameters and other factors that are related to this range. For thinner fibers, resolutions can improved. As from the observation seen in Figs. Figs. \ref{fig5} (c) and (d),  the present technique has been able to predict the diameter of TFs with a variation as low as $\sim$150 nm. Note that the accuracy in measuring the TF diameter depends entirely on the PD's accuracy and the light source's stability. Light source can be split using a 50:50 beam splitter. One would directly monitor the fluctuations from the source at the time of measurement, serving as reference and other would be passed through TF. Based on the fluctuations in the reference channel, improvements can be made in the readings obtained at detector end. This approach can minimize the error and improve signal to noise ratio to a large extent. We would also like to point out that range of  TF diameter probed in this case ranged were from 2 $\mu$m to 11 $\mu$m. To avoid mismatch in continuity of FESEM frames of TF, we choose magnification of 2000, a scale that covers the probing sizes effectively. To maintain continuity in the FESEM readings, a marker would be  identified in each frame and that marker would serve as reference in the following FESEM frame. Within the frame, contrast between the adjacent pixels along the lines of measurement was used for determining the periphery of the fiber. These pixel to pixel distances were used as diameter readings. We used ImageJ software for these readings. The experimental results very clearly and accurately predicted the relative diameter of the TF with respect to the waist region, highlighting the irregularities in the diameter profile. The measured results agree with the simulation results, within experimental error. We also believe this method can produce much better results with TFs fabricated from more conventional methods such as heat and pull technique \cite{038,039}.
	
Note that Whispering Gallery Modes (WGM) are dependent on wavelength of the source and size of the resonator. However, since we have selected a single wavelength for our simulations and experiments, we can safely say that if we choose the thickness of the PF carefully, we can avoid WGMs. Moreover, we are focused on relative scattering loss study, any resonance effects will also be proportional to intensity itself.

\section{Conclusion}
In summary, we experimentally demonstrated the in-situ characterization of optical micro/nano fibers (MNFs). We measured the scattering loss along the MNF axis at various diameters by mounting it on the probe fiber. The diameter profile of the MNF was inferred from the measured scattering loss and correlated with its surface morphology measurement. The experimental results were reproduced within experimental ambiguity with simulation-predicted results. The present technique may open new possibilities in the sensing field and become a standard tool. This work demonstrates an effective, low-cost, and non-destructive method for in-situ characterization of fabricated micro/nano fibers (OMNFs). It can detect and determine the irregularities on the surface of the OMNFs. It can also be used to quantify the local evanescent field. Detecting such local points can improve studies that are carried out using these fields in various sensing and related study domains. Upon calibration, this method can also determine the local diameter with great accuracy. It is an all-round simplified technique for various purposes, as pointed out. Another advantage of this method is re-usability of the fabricated sample after optical characterization. It is simple to implement and can be accessed by all domains of researchers.

\begin{acknowledgments}
SS acknowledges funding support for the Chanakya-PG fellowship from the National Mission on Interdisciplinary Cyber-Physical Systems, of the Department of Science and Technology, Govt. of India through the I-HUB Quantum Technology Foundation (File No. I-HUB/PGF/2022-23/01). RM acknowledges the University Grants Commission (UGC) for the financial support (Ref. No.:1412/CSIR-UGC NET June 2019). RRY acknowledges the financial support from the Institute of Eminence (IoE) grant at the University of Hyderabad (File No. RC2-21-019) and the Scheme for Transformational and Advanced Research in Sciences (STARS) from the Ministry of Human Resource Development (MHRD) (File No. STARS/APR2019/PS/271/FS). 
\end{acknowledgments}
\section*{Disclosures}
The authors declare no conflicts of interest.
\section*{Author Contributions}
SS and EB have contributed equally in this work.
\section*{Data Availability Statement}
Data underlying the results presented in this paper are not publicly available at this time but may be obtained from the authors upon reasonable request.

\appendix

\bibliography{./RefEXP}

\end{document}